\documentclass[runningheads]{llncs}
\usepackage{graphicx}
\usepackage{float}
\usepackage{todonotes}

\begin{document}
\title{Ontology Extraction and Usage in the Scholarly Knowledge Domain\thanks{We would like to thank Springer Nature for partially funding this research.}}
%
%\titlerunning{Abbreviated paper title}
% If the paper title is too long for the running head, you can set
% an abbreviated paper title here
%
\author{Angelo A. Salatino\orcidID{0000-0002-4763-3943} \and
Francesco Osborne\orcidID{0000-0001-6557-3131} \and
Enrico Motta\orcidID{0000-0003-0015-1952}}
\authorrunning{A. Salatino et al.}
% First names are abbreviated in the running head.
% If there are more than two authors, 'et al.' is used.
%
\institute{Knowledge Media Institute, The Open University, Milton Keynes, United Kingdom\\
\email{\{angelo.salatino,francesco.osborne,enrico.motta\}@open.ac.uk}}
\maketitle              % typeset the header of the contribution
\begin{abstract}
Ontologies of research areas have been proven to be useful resources for analysing and making sense of scholarly data. In this chapter, we present the Computer Science Ontology (CSO), which is the largest ontology of research areas in the field, and discuss a number of applications that build on CSO to support high-level tasks, such as topic classification, metadata extraction, and recommendation of books.

\end{abstract}
\section{Introduction}

Ontologies, as formal specifications of concepts and relations in specific domains, have become a standard solution to represent domain knowledge, integrate data from different sources, and support a variety of semantic applications \cite{dinoia2018,kishore2007}. In the field of scholarly knowledge, ontologies are used to facilitate the integration of large datasets of research data \cite{livingston2015}, the exploration of the academic landscape \cite{osborne2013,kirrane2019decade}, information extraction from scientific articles \cite{fathalla2017}, and so on. More specifically, ontologies representing research topics and describing their relationships, have been employed in several tasks, such as making sense of research dynamics \cite{osborne2013}, classifying research publications \cite{salatino2019c}, characterising \cite{bettencourt2009} and identifying \cite{osborne2014} research communities, studying the origin of research topics \cite{salatino2017topics}, and forecasting research trends \cite{salatino2018}.
%\textcolor{red}{Ontologies have proved to be powerful solutions to represent domain knowledge, integrate data from different sources, and support a variety of semantic applications [1–5]. In the scholarly domain, ontologies are often used to facilitate the integration of large datasets of research data [6], the exploration of the academic landscape [7], information extraction from scientific articles [8], and so on. Specifically, ontologies that describe research topics and their relationships are invaluable tools for helping to make sense of research dynamics [7], to classify publications [3], to characterise [9] and identify [10] research communities, and to forecast research trends [11].}
However, not every domain of science has an ontology that comprehensively describes all research concepts and their relations. In addition, ontologies describing research topics are typically manually crafted by domain experts which is a very time consuming process. Therefore, they usually evolve slow and become quickly outdated. A further issue is that many of these ontologies tend to be very coarse-grained, lacking the right depth that would allow them to comprehensively describe the area. All these limitations hinder the adoption of semantic technologies in several fields of science.

The question we are left with is: how do we effectively extract comprehensive, fine-grained, and  up-to-date ontologies of research topics that can support a range of intelligent services?

\begin{figure}[H]
\center{\includegraphics[width=0.6\textwidth]
{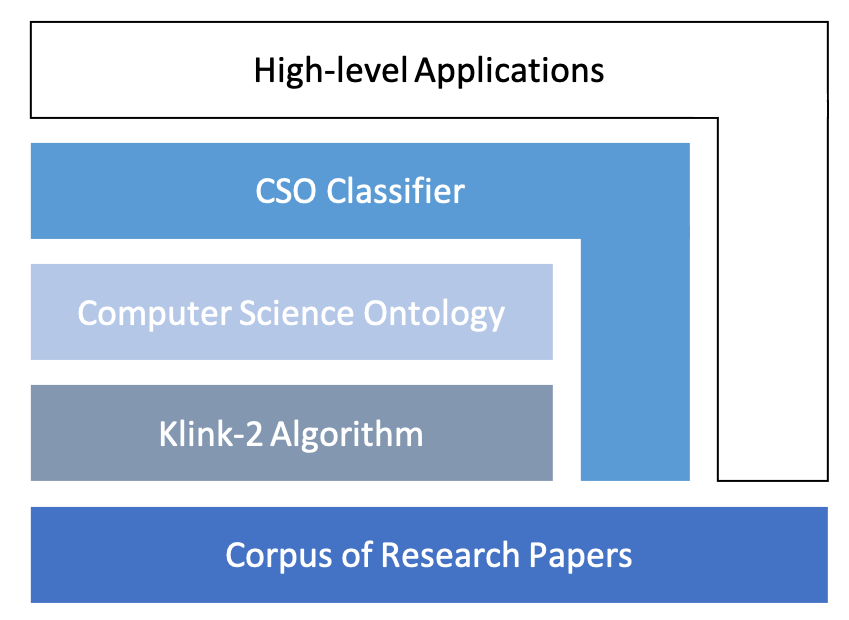}}
\caption{\label{fig:stack} The Computer Science Ontology Framework.}
\end{figure}

In this chapter, we present the Computer Science Ontology (CSO) Framework \cite{salatino2018computer,salatino2019b}, which is a conceptual framework for generating a large scale ontology of Computer Science. This solution has been used to support a variety of high-level tasks, such as  (i) categorising proceedings in digital libraries, (ii) enhancing semantically the metadata of scientific publications, (iii) generating recommendations, (iv) producing smart analytics, (v) detecting research trends, and others \cite{osborne2016,salatino2018computer}.
%and exploiting the Computer Science Ontology. %This is a general architecture that could be implemented in other domain, 
%that characterises the design, extraction and reasoning with
%ontologies in the scholarly domain.
Figure \ref{fig:stack} shows the architecture of the framework, where each layer exploits the underneath layers.

The first layer is the corpus of research papers. This represent the core input data from which we want to extract a granular and data-driven ontology of research areas.
%describing most branches of science. %In this paper, we will use the Rexplore dataset \cite{osborne2013} consisting of 16M scientific articles in the field of Computer Science.

Sitting on top of the data layer, we find the Klink-2 algorithm \cite{osborne2015} that generates the large-scale ontology of research topics from publication metadata.
The Computer Science Ontology (CSO), which represents the third layer, is a large-scale, granular, and automatically generated ontology of research areas. The current version includes about 14K topics and 163K semantic relationships. 
At the fourth layer, we find the CSO Classifier \cite{salatino2019c}, a tool for automatically classifying research papers according to the topics available in the Computer Science Ontology. This system enables users to represent scientific publications in terms of CSO concepts and allows all relevant stakeholders to develop a multitude of relevant smart functionalities.

The rest of the chapter is organised as follows. In Section 2, we review the literature regarding topic detection in research papers, pointing out the existing gap. In Section 3, we describe the Klink-2 algorithm for generating the ontology, and in Sections 4 we discuss CSO with its data model. In Section 5 we describe the CSO Classifier and in Section 6 we show in more detail some high-level applications that allowed us to perform smart analytics on scholarly data. Finally, in Section 7 we summarise the main conclusions and outline future directions of research.

\section{Literature Review}
Some fields of research are comprehensively described by ontologies of research areas, e.g., MeSH in Biology and PhySH in Physics. These ontologies can provide support toward a multitude of tasks, such as integrating heterogeneous datasets \cite{livingston2015}, assisting users in exploring digital libraries \cite{osborne2013}, producing scholarly analytics \cite{boyack2011}, and forecasting research dynamics \cite{salatino2018,osborne2016}. In this section, we will review the current state of the art with regards to developing and using ontologies of research areas. In particular, we will first provide an overview of some of most well-known ontologies of research areas, then we will discuss current approaches for the generation of these ontologies, and finally, we will describe the approaches that take advantage of such ontologies to perform several high-level tasks.

\subsection{Ontologies of research areas}
In literature, we can find different ontologies of research areas, which are scoped to a particular branch of science. In the field of Computer Science, the most well-known taxonomy is the ACM Computing Classification System\footnote{ACM Computing Classif. System - \url{https://www.acm.org/publications/class-2012}}, developed and maintained by the Association for Computing Machinery (ACM). This taxonomy contains around 2K concepts and it is manually curated. These characteristics represent a limitation as its representation of the field lacks both depth and breadth, and its curation process is slow and expensive.

In the field of Medicine, the most popular solution is the Medical Subject Heading (MeSH)\footnote{Medical Subject Heading - \url{https://www.ncbi.nlm.nih.gov/mesh}} maintained by the National Library of Medicine of the United States. This taxonomy is constantly updated by collecting new terms as they appear in the scientific literature.

The Physics Subject Headings (PhySH)\footnote{Physics Subject Headings - \url{https://physh.aps.org/}} is the standard solution in the field of Physics and Astronomy. It is developed by the American Physical Society (APS) and it is constantly updated with the support of authors, reviewers, editors and organisers of scientific conferences.

In the field of Mathematics there is the Mathematics Subject Classification (MSC)\footnote{Mathematics Subject Classification - \url{https://mathscinet.ams.org/msc}} which is maintained by Mathematical Reviews and zbMATH. In the field of Economics we can find the JEL classification\footnote{Journal of Economic Literature - \url{https://www.aeaweb.org/econlit/jelCodes.php}}, created by the Journal of Economic Literature of the American Economic Association, and the STW Thesaurus for Economics\footnote{STW Thesaurus for Economics - \url{http://zbw.eu/stw/}} developed by ZBW - Leibniz Information Centre for Economics.

The ontologies mentioned above can comprehensively represent specific areas of science. However, in literature we can also find more broad ontologies covering a multitude of fields.
The most popular ones are the Library of Congress Classification and the Dewey Decimal Classification which encompass many areas of science. Indeed, these two schemes are employed to classify books within large academic libraries, globally.

Another ontology of research areas in the Fields of Research (FoR) which is developed by the Australian Bureau of Statistics, New Zealand Ministry of Business, Innovation, and Employment, and other partners. This scheme covers many areas of science and indeed it is currently adopted by Dimensions.ai\footnote{Dimensions.ai - \url{https://www.dimensions.ai}}, a company that provides commercial solutions to support users in exploring the research landscape.

A common issue of these ontologies is that, being manually crafted and maintained by domain experts, they tend to evolve relatively slow and  became quickly outdated. To keep-up with the pace of the constant evolution of the research landscape, some institutions (e.g., the American Physical Society) are crowd-sourcing their classification scheme. However, crowd-sourcing strategies suffer from limitations, such as trust and reliability \cite{clough2013}. Indeed, those institutions need to entrust a committee to moderate such amendments to the classification scheme. %In the next section we will show some progress made toward the process of automatic generation of ontologies of research area.

%Since generating these knowledge bases manually requires a large number of experts and is an expensive and lengthy process, in the last years we saw the emergence of several methods for producing them (semi-) automatically from a set of relevant documents. In this section, we will first focus on current available ontologies of research topics and then discuss the approaches to automatically generate them.

\subsection{Automatic extraction of ontologies from scholarly data}
A new strategy which is becoming increasingly popular is the automatic or semi-automatic generation of ontologies of research topics using data-driven methodologies. In the state of the art we can find a variety of approaches that allow us to learn ontologies using clustering techniques, natural language processing, statistical methods and others.

TaxGen \cite{muller1999} is an approach for the automatic generation of taxonomies from a corpus using both hierarchical agglomerative clustering algorithm and text mining techniques. The algorithm, in a bottom-up fashion, first identifies the bottom clusters by observing the linguistic features in the documents, such as co-occurrences of words, domain terms, names of people, organisations and other significant words from text. Then the clusters are merged creating higher-level clusters, which form the hierarchy.

Text2Onto \cite{cimiano2005} is another approach for learning ontologies from a collection of documents. This approach identifies sub-/superclass hierarchies, synonyms, and other linguistic features through the application of natural language processing techniques on the sentence structure, where phrases like “such as…” and “and other…” imply a hierarchy between terms. This method presents some similarities with the Klink-2 algorithm \cite{osborne2015}, which we will describe later, but requires the full text of documents.

Sanderson et al. \cite{sanderson1999} developed an approach for automatically deriving a hierarchical organisation of concepts from a set of documents without use of training data. Their approach computes the conditional probability for a keyword to be associated with another based on their co-occurrence. Given a pair of keywords, this system tries to understand whether there is a subsumption relationship between them, according to certain heuristics.

In literature, we can also find semi-automatic approaches that take advantage of external knowledge, either from pre-existing taxonomies or sourced by the community. For instance, Shen et al. \cite{shen2018} developed the Fields of Study (FoS) taxonomy, currently in use within Microsoft Academic, in which the first two levels are hand-crafted. Their approach is based on a variation of \cite{sanderson1999} and automatically infers topics derived from Wikipedia. However, considering only Wikipedia is a limitation as many research topics are not described there. Conversely, Klink-2 considers both academic publications and external sources. Another approach from Wohlgenannt et al. \cite{wohlgenannt2012}, combines human effort and machine computation by crowd-sourcing the evaluation of an automatically generated ontology with the aim of dynamically validating the extracted relations. Klink-2 can benefit these systems by generating an accurate, large-scale and up-to-date topic network.

%\subsection{Reasoning with scholarly ontologies}
%To better \todo{Not sure if this area is actually useful at all. I just wanted to add some approaches in Scholarly domain that have been successful with ontologies. But later we show our high level application, maybe we can remove it.} appreciate the value of using ontologies of research areas when performing analyses on scholarly data, we need to look at some successful stories in literature. Indeed, the state of the art provides enough evidence on how the adoption of ontologies of research area can be beneficial to improve the performance of a multitude of tasks.

%Onhiwa et al. \cite{ohniwa2010} offered an approach for categorising Medicine literature described through the MeSH subjects.
%In particular, they create a scientific concept network in which each node represents a MeSH subject, which refers to a specific topic in Medicine. Two nodes are then linked if the two corresponding codes co-occur together in at least one paper.
%%Since generating these knowledge bases manually requires a large number of experts and is an expensive and lengthy process, in the last years we saw the emergence of several methods for producing them (semi-) automatically from a set of relevant documents. In this section, we will first focus on current available ontologies of research topics and then discuss the approaches to automatically generate them.

\section{Ontology Generation}\label{sec:onto-generation}
%For an ontology of research areas it is essential to describe the structures of branches and sub-branches within a discipline.
An essential characteristic for ontologies of research areas is the hierarchical description of branches and sub-branches for a given discipline. 
Indeed, almost all ontologies mentioned above, e.g. MeSH, PhySH, ACM-CSS, are structured like a taxonomy: from the most generic to the most specific topics. In cases when there are different labels associated to the same topic, for instance for acronyms or synonyms, the ontology needs to be modelled in a way that can contain information about equivalences.

%Hence, it is crucial that an ontology of research areas contains hierarchical relationships, and in case there are several labels describing the same concept these ontologies should contain information about equivalences.
To this end, we designed the Klink-2 algorithm \cite{osborne2015} that from a large corpus of scholarly publications is able to automatically generate ontologies of research areas, describing both hierarchical and equivalence relationships between research topics. %Specifically, Klink-2 ingests a large corpus of scholarly publications and returns an ontology describing the both hierarchical and equivalence relationships between research topics.
%In order to extract ontologies of research areas in an automatic way we designed the Klink-2 algorithm \cite{osborne2015}. Klink-2 is able ingest large corpus of scholarly publications and return an ontology describing the semantic relationships between research topics. 

The algorithm starts by comparing each keyword in input to all the other keywords with which it shares at least \textit{n} co-occurrences. In particular, it infers the semantic relationship between topics \textit{x} and \textit{y} using three metrics: i) $H_R (x,y)$, which uses a semantic variation of the subsumption method \cite{sanderson1999} for measuring the intensity of a hierarchical relationship; ii) $T_R (x,y)$, which uses temporal information to estimate the existence of a hierarchical relationship; and iii) $S_R (x,y)$, which estimates the similarity between two topics. The first two metrics are used to indicate whether a topic is super-area of another one (\textit{superTopicOf}) or that the research outputs of one topic contributes to research of the other (\textit{contributesTo}). The last metric is used to infer equivalence relationships between topics (\textit{relatedEquivalent}). 

$H_R (x,y)$ quantifies the hierarchical relationship between x and y according to the following formula:
\begin{equation}
\label{eq:hr}
H_{R}(x, y)=\left(\frac{I_{R}(x, y)}{I_{R}(x, x)}-\frac{I_{R}(y, x)}{I_{R}(y, y)}\right) \cdot c_{R}(x, y) \cdot n(x, y)
\end{equation}

where $I_R (x,y)$  is the number of elements associated with both \textit{x} and \textit{y} according to relation \textit{R} (e.g., number of co-occurrences in research papers), ($\frac{I_{R}(x, y)}{I_{R}(x, x)}$) is the conditional probability that an element associated with keyword \textit{x} will be associated also with keyword \textit{y}, $n (x,y)$ defines the string similarity between the two topics using the normalised Levenshtein distance, and finally $c_R (x,y)$ measures how similar are the distributions of topics with which both topic \textit{x} and \textit{y} are co-occurring, using cosine similarity.

$T_R (x,y)$ is a temporal version of $H_R (x,y)$, which weighs more the information associated with the first years of \textit{x}. It is useful to detect the cases in which the relationship between two terms fades because their association has become implicit (e.g., Artificial Intelligence and Machine Learning). $T_R (x,y)$ is calculated using a variation of Eq. \ref{eq:hr} in which $I_R (x,y)$  is computed by weighting the intensity of the relationships in each year according to the distance from the debut of x. The weight is computed as $w(year, x)= (year - debut(x) +1) -\gamma$, with $\gamma>0$ ($\gamma=2$ in the prototype).
Finally, $S_R (x,y)$ is used to assess the similarity of two terms and is computed according to the following formula:

\begin{equation}
S_{R}(x, y)=\frac{c_{R}(x, y)}{\max \left(c_{R}^{super}(x, y), c_{R}^{sib}(x, y)\right)+1}
\end{equation}
 
where $c_{R}^{super}(x, y)$ is the cosine similarity of the super topics of the two terms in the taxonomy produced by previous iteration, and $c_{R}^{sib}(x, y)$ is the cosine similarity of their siblings.
A hierarchical relationship between two topics is inferred when a sufficient number of hierarchical indicators are above a threshold. An analysis of the precision/recall trade-off associated with different thresholds is available in \cite{osborne2015}. The nature of the inferred relationship is assessed by Klink-2 using a rule-based approach. In brief, if \textit{x} is older, associated with more entities, and the $T_R (x,y)$ indicators score higher, Klink-2 will infer a \textit{superTopicOf} relationship, otherwise a \textit{contributesTo} one. Then, Klink-2 removes loops in the topic network, merges keywords linked by a \textit{relatedEquivalent} relationship, and splits ambiguous keywords associated to multiple meanings (e.g., “Java”). The keywords produced in this step are added to the initial set of keywords to be further analysed in the next iteration and the while-loop is re-executed until there are no more keywords to be processed. 

Klink-2 filters the keywords considered too generic or not academic according to a set of heuristics that take in consideration the frequency of a keyword in various online sources and distribution of its co-occurrences \cite{osborne2012mining}. Finally it generates the RDF triples describing the ontology. For a more comprehensive explanation of Klink-2, we refer the reader to Osborne at al. \cite{osborne2015}.

\section{Computer Science Ontology}
The Computer Science Ontology (CSO) is a large-scale, granular, and automatically generated ontology of research areas. It was generated by running the Klink-2 algorithm on the Rexplore dataset \cite{osborne2013} containing 16 million publications  in the field of Computer Science.

%, to automatically generate a large-scale ontology of research areas in this field: the Computer Science Ontology (CSO).
%The Computer Science Ontology is a large-scale ontology of research areas that was automatically generated using the Klink-2 algorithm \cite{osborne2015} on a dataset of about 16 million publications, mainly in the field of Computer Science \cite{salatino2018b}. In the rest of the paper, we will refer to this corpus as the Rexplore dataset \cite{osborne2013}.

Currently CSO includes about 14K topics and 163K semantic relationships. The main root is Computer Science; however, the ontology includes also a few secondary roots, such as Linguistics, Geometry, Semantics, and so on.
The CSO data model\footnote{CSO data model - \url{https://cso.kmi.open.ac.uk/schema/cso}} is an extension of SKOS\footnote{SKOS Simple Knowledge Organization System - \url{http://www.w3.org/2004/02/skos}.}  and it includes eight semantic relations:
\begin{itemize}
    \item \textit{relatedEquivalent}, which is a subproperty of skos:related, indicates that two topics can be treated as equivalent for the purpose of exploring research data (e.g., Ontology Matching and Ontology Mapping). 
    \item \textit{superTopicOf}, which is a subproperty of skos:narrower, indicates that a topic is a super-area of another one (e.g., Semantic Web is a super-area of Linked Data). The inverse of this relationship is subTopicOf. 
    \item \textit{contributesTo}, which indicates that the research output of one topic contributes to another. For instance, research in Ontology Engineering contributes to Semantic Web, but arguably Ontology Engineering is not a sub-area of Semantic Web, since there is plenty of research in Ontology Engineering outside the Semantic Web area.
    \item \textit{owl:sameAs}, which is used for mapping CSO topics to equivaled entities in other knowledge graphs (DBpedia\footnote{DBpedia - \url{https://wiki.dbpedia.org}}, Freebase\footnote{Freebase - \url{https://en.wikipedia.org/wiki/Freebase}}, Wikidata\footnote{Wikidata - \url{https://www.wikidata.org}}, YAGO\footnote{Yago - \url{https://github.com/yago-naga/yago3}}, and Cyc\footnote{Cyc - \url{https://www.cyc.com}}).
    \item \textit{schema:relatedLink}, which links CSO concepts to relevant web pages that either describe the research topics (Wikipedia articles) or provide additional information about the research domains (Microsoft Academic).
    \item \textit{preferentialEquivalent}, which is used to state the main label for topics belonging to a cluster of relatedEquivalent. For instance, the topics Ontology Matching and Ontology Alignment both have their preferentialEquivalent set to Ontology Matching. Similarly to relatedEquivalent, in our data model we defined preferentialEquivalent as a subproperty of skos:related.
    \item \textit{rdf:type}, this relation is used to state that a resource is an instance of a class. For example, a resource in our ontology is an instance of Topic, which is a subclass of skos:Concept. 
    \item \textit{rdfs:label}, this relation is used to provide a human-readable version of a resource’s name. 
\end{itemize}

In the previous section, we discussed how the Klink algorithm infers the first three of these relationships: \textit{relatedEquivalent}, \textit{superTopicOf}, and \textit{contributesTo}. The rest of the relationships are also automatically generated.
In particular, the \textit{rdf:type} and \textit{rdfs:label} relations respectively identify all topic entities and their labels. We generated the \textit{preferentialEquivalent} choosing, within a cluster of topics linked by a \textit{relatedEquivalent}, the label associated with most articles in the source corpus \cite{osborne2013}.
To generate the \textit{owl:sameAs} relationships, linking CSO concepts to equivalent entities in the other KBs, we identified the DBpedia entities corresponding to CSO topics by exploiting the DBpedia Spotlight API \cite{mendes2011}. Then we extracted the links from DBpedia to other KBs in the Linked Open Data (LOD) cloud by using the DBpedia SPARQL endpoint. 

We also generated \textit{schema:relatedLink} relations toward external web pages containing further information about the research topic. In particular, the links between CSO concepts and Wikipedia articles were extracted from the DBpedia. We also mapped the CSO topics to the Fields of Study (FoS) concepts reseased by Microsoft Academic. More details regarding the alignment between CSO and other knowledge bases are avaliable in Salatino et al. \cite{salatino2019b}.

\begin{figure}[t]
\center{\includegraphics[width=\textwidth]
{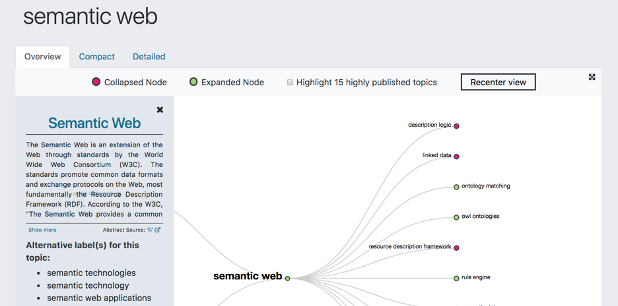}}
\caption{\label{fig:cso-portal} Overview of the resource page related to the topic in a new ``semantic web''.}
\end{figure}

To facilitate the uptake of CSO we have developed the CSO Portal\footnote{CSO Portal - \url{https://cso.kmi.open.ac.uk}} (see Fig. \ref{fig:cso-portal}), a web application that enables users to browse,  download – in various formats\footnote{This ontology is licensed under a Creative Commons Attribution 4.0 International License (CC BY 4.0) - \url{https://creativecommons.org/licenses/by/4.0.}.}, e.g. N-Triples, OWL, TTL and CSV – and provide granular feedback on CSO at different levels. %The feedback from the community will then be used to generate new versions of CSO.

\section{CSO Classifier}
\begin{figure}[!htb]
\center{\includegraphics[width=\textwidth]
{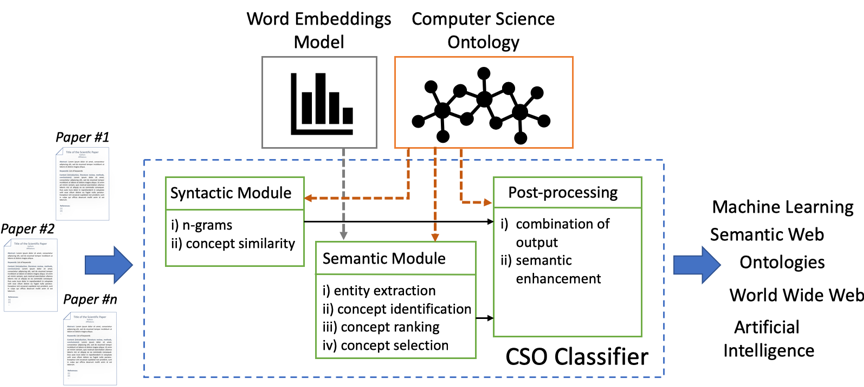}}
\caption{\label{fig:cso-workflow} Workflow of the CSO Classifier.}
\end{figure}

In order to facilitate users in integrating CSO in their pipelines,  we developed the CSO Classifier \cite{salatino2019c}, a tool that automatically annotates documents according to CSO. This application takes in input the metadata associated with a research paper (title, abstract, and keywords) and returns a selection of research concepts drawn from CSO. Figure \ref{fig:cso-workflow} displays its workflow.

The CSO Classifier works in three steps. First, it finds all topics in the ontology that are explicitly mentioned in the paper (syntactic module). Then it identifies further semantically related topics by means of part-of-speech tagging and word embeddings (semantic module). Finally, it enriches this set by including the super-areas of these topics according to CSO.

In particular, the \textit{syntactic module} removes English stop words and collects unigrams, bigrams, and trigrams. Then, for each n-gram, it computes the Levenshtein similarity with the labels of the topics in CSO. Finally, it returns all research topics whose labels have similarity to one of the n-grams, which is equal to or higher than a threshold.

The \textit{semantic module} takes advantage of a pre-trained word embedding model which captures semantic properties of words \cite{mikolov2013}. We trained this Word2Vec model using titles and abstracts of 4,654,062 English publications in the field of Computer Science from Microsoft Academic Graph, which is an heterogeneous graph containing scientific publication records, citation relationships, authors, institutions, journals, conferences, and fields of study. We pre-processed this data by replacing spaces with underscores in all n-grams matching the CSO topic labels (e.g., “semantic web” became “semantic\_web”) as well as for frequent bigrams and trigrams (e.g., “highest\_accuracies”, “highly\_cited\_journals”). The latter were identified by analysing collocations, i.e. combinations of words that co-occur together \cite{mikolov2013}. This solution allows the CSO Classifier to better disambiguate concepts and treat terms such as “deep\_learning” and “e-learning” as completely different words.

Specifically, to compute the semantic similarity between the terms in the document and the CSO concepts, the semantic module uses part-of-speech tagging to identify candidate terms composed by a combination of nouns and adjectives and decomposes them into unigrams, bigrams, and trigrams. For each n-gram, it retrieves its most similar words from the Word2Vec model. For this task, the n-gram tokens are initially glued with an underscore, creating one single word, e.g., “semantic\_web”. If this word is not available within the model vocabulary, the classifier uses the average of the embedding vectors of all its tokens. Then, it computes the relevance score for each topic in the ontology as the product between the number of times it was identified in those n-grams (frequency) and the number of unique n-grams that led to it (diversity). Finally, it uses the elbow method \cite{satopaa2011} for selecting the set of most relevant topics. 
The CSO Classifier aggregates the topics returned by the two modules and enriches them by inferring the list of all their super topics, exploiting the \textit{superTopicOf} relationship within CSO. For instance, given the topic “Neural Networks”, it will infer “Machine Learning”, “Artificial Intelligence”, and “Computer Science”. This feature allows us to capture both high-level fields and very granular research areas, in order to generate a comprehensive representation of the classified papers.

%The Python implementation of the latest version of the CSO Classifier is available at \url{https://github.com/angelosalatino/cso-classifier}.
The latest release of the CSO Classifier can be installed via \textit{pip} from PyPI: pip install cso-classifier; or it simply downloading it from \url{https://github.com/angelosalatino/cso-classifier}.

\section{High-level applications}
This section describes some high-level applications that take advantage of CSO for supporting users in exploring, analysing, and making sense of large corpora of research publications.

\subsection{Exploring and making sense of scholarly data}
Rexplore is a system to support users in exploring and making sense of scholarly data \cite{osborne2013}.
%leveraging novel solutions in large-scale data mining, semantic technologies and visual analytics.
%Rexplore \cite{osborne2013}  
It uses CSO to characterise research papers, authors, and organisations according to research topics. 
%An interesting feature available in Rexplore is that it can plot the collaboration graph of researchers based on their topic similarity, reflecting how similar two authors are with respect to their research areas.
An interesting feature available in Rexplore is that it can plot a graph of researchers based on their topic similarity, reflecting how similar two authors are with respect to their research areas.
Rexplore allows users to detect and make sense of important trends in research, such as significant migrations of researchers from one area to another, the emergence of new topics, the evolution of communities within a particular area.
It also provides powerful query/search facilities, supporting complex multidimensional queries that can include logical connectives, such as retrieving career-young authors who have worked in both \textit{Semantic Web} and \textit{Social Networks}, and have published at the International Semantic Web Conference.

The Rexplore system was shown to be able to support users in performing specific tasks more effectively than Microsoft Academic Search (MAS), thanks to its organic representation of research topics \cite{osborne2013}.

\subsection{Automatic classification of conference proceedings}
%Classifying conference proceedings and other editorial products according to their relevant research areas facilitates their discovery and allows editors to take informed decisions on how to better marketise them.

The Smart Topic Miner (STM) \cite{salatino2019} is a web application that supports the Springer Nature editorial team in classifying editorial products according to a taxonomy of research topics drawn both from CSO and the Product Market Codes (PMC), Springer Nature’s own editorial classification system. 
%This information is then used for: i) classifying proceedings in digital and physical libraries; ii) enhancing semantically the metadata associated with publications and consequently improving the discoverability of the proceedings; and iii) detecting promising emerging research areas that may deserve more attention from the publisher. 
STM takes as input the metadata associated with the proceedings of a conference (titles, abstracts and author-provided keywords for each paper in the proceedings) and returns the set of relevant CSO topics and PMCs as output.% (as showed in Fig. \ref{fig:stm}). 

STM uses the CSO Classifier to annotate each paper with the topics from CSO. Then it groups and ranks the topics according to the number of papers addressing them. Finally, it infers the relevant PMCs, using the mapping between the CSO ontology and PMC. The editors then review the CSO topics and the PMC categories and submit these annotations to the Springer Nature production system. This outcome is displayed in the Springer Nature’s digital library: SpringerLink; and included in the ONIX\footnote{ONIX for Books - \url{https://bisg.org/page/ONIXforBooks}} metadata feeds, delivered to various libraries and bookshops. 

We released STM back in 2016, and since then it has been routinely used by the editorial team to annotate all book series covering conference proceedings in Computer Science, including LNCS, LNBIP, CCIS, IFIP-AICT and LNICST, for an amount of 800 volumes per year.
STM has the advantage of halving the time required for classifying proceedings and it also reduced the complexity of the task, which  was traditionally carried out by senior editors but now performed by junior editors. Its adoption produced a significant increment of the discoverability of relevant publications on SpringerLink, resulting in about 9 million additional downloads over the last three years. A demo of STM is available at \url{http://stm-demo.kmi.open.ac.uk/}. 

%STM was introduced in 2016 [3] and has since been used routinely by the editorial team to annotate all book series covering conference proceedings in Computer Science, including LNCS, LNBIP, CCIS, IFIP-AICT and LNICST, for a total of about 800 volumes each year. During this period, the adoption of STM has halved the time needed for classifying proceedings from 20-30 to 10-15 minutes. In addition, STM also provided the additional important benefit of reducing the complexity of this task, which traditionally has been performed by Senior Editors. Indeed, thanks to the introduction of STM in the editorial workflow, it has now become possible for this task to be carried out by junior editors, ultimately achieving an overall 75\% cost reduction. 

% \begin{figure}[t]
% \center{\includegraphics[width=\textwidth]
% {figures/STM_1.png}}

% \caption{\label{fig:stm}A snapshot of the Smart Topic Miner interface.}
% \end{figure}

\subsection{Book recommendation}
%A second application we developed for the Springer Nature editorial team is 
The Smart Book Recommender (SBR) \cite{thanapalasingam2018} is an ontology-based recommender system that supports the Springer Nature editorial team in promoting their publications at Computer Science venues. It takes as input the proceedings of a conference and returns books, journals and other proceedings that are likely be of interest for its attendees. %(see Fig. \ref{fig:sbr}).

%It does so by computing the similarity of SN editorial products over the vectors of semantic topics returned by the CSO Classifier.

%The Smart Book Recommender (SBR) is an ontology-based recommender system that supports the Springer Nature editorial team in promoting their publications at Computer Science venues. It takes as input the proceedings of a conference and suggests books, journals, and other conference proceedings which are likely to be relevant to the attendees of the conference in question. 
SBR uses the CSO Classifier to represent more than 27K books and 320 journals according to their distribution of topics. Then, it identifies the most relevant editorial products, computing the similarity between the topical representation of the input conferences and other products available in the system. SBR also exploits the CSO topic taxonomy to graphically represent and compare conferences and books, allowing users to understand the rationale behind its recommendations. A demo of SBR is available at \url{http://rexplore.kmi.open.ac.uk/SBR-demo}.

% \begin{figure}[t]
% \center{\includegraphics[width=\textwidth]
% {figures/SBR_1.png}}

% \caption{\label{fig:sbr}A snapshot of the Smart Book Recommender interface.}
% \end{figure}

\subsection{Forecasting research topics}

Understanding and reacting timely to new developments in the research landscape is critical for a variety of stakeholders, such as funding bodies, academic publishers, companies and others. Augur \cite{salatino2018} is a novel approach which uses CSO for anticipate the emergence of new research topics. Specifically, Augur analyses \textit{topic networks}, i.e. collaboration networks between research communities associated with specific research areas, and identifies clusters associated with a significant increase in the pace of collaboration. Over these networks, Augur applies a novel clustering algorithm called the Advanced Clique Percolation Method (ACPM). The resulting clusters of topics indicate the areas of the network that are nurturing new research areas. 
Augur uses CSO for creating semantically-enhanced topic networks describing the collaboration between research topics over time. The evaluation of Augur proved that semantically enriching topics networks with CSO yields more than 30\% increase of f-measure on the task of predicting the emergence of new research areas.
%, compared to the solution in which keywords represent research topics.
Further details of Augur and its evaluation are available in Salatino et al. \cite{salatino2018}.

\subsection{Systematic literature reviews}
%Systematic reviews aim to find as much as possible of the research relevant to the particular research questions, and use explicit methods to identify what can reliably be said on the basis of these studies.
The aim of systematic reviews (SRs) is to find all the evidence relevant to a particular research question, and identify what can be said based on those findings.
%The aim of systematic reviews is to collect all the evidence to the particular research questions, and rigorously synthesising those findings.
Typically, systematic reviews require domain experts to collect, annotate and synthesise hundreds of papers manually, using a well-defined methodology meant to mitigate the risks of biases and ensure repeatability for later updates. This task becomes extremely hard when investigating large numbers of papers (e.g. hundreds
of thousands). The Expert-Driven Automatic Methodology (EDAM) \cite{osborne2019} was developed for  reducing the amount of tedious manual tasks involved in SRs while taking advantage of the value provided by human expertise.
%Systematic reviews are means for collecting and synthesising evidence from the identification and analysis of relevant studies from multiple sources. To this aim, they use a well-defined methodology meant to mitigate the risks of biases and ensure repeatability for later updates. 
%which involves both automatic techniques and human experts;
%To support users in this task, we developed the Expert-Driven Automatic Methodology (EDAM) \cite{osborne2019} that reduces the amount of manual tedious tasks involved in SRs while taking advantage of the value provided by human expertise.
%EDAM \cite{osborne2019} is an expert-driven automatic methodology for creating systematic reviews that limits the amount of tedious tasks that have to be performed by human experts. 
%Typically, systematic reviews require domain experts to annotate hundreds of papers manually. 
In particular, EDAM is able to i) characterise the area of interest using an ontology of topics, ii) ask domain experts to refine such an ontology, and iii) take advantage of this knowledge base for classifying relevant papers and producing useful analytics. 

This approach uses the CSO Classifier to classify all research papers using title, abstract, and keywords. For a given topic, it then categorises all papers that were annotated with that specific topic, as well as all its \textit{relatedEquivalent} and all its sub-branches, using the \textit{superTopicOf} relation within CSO. 

We evaluated the ability of EDAM to correctly discriminate between different topics in the field of Software Architecture by classifying a set of randomly-selected papers both with EDAM and with six human experts. We compared the annotation produced by both human experts and EDAM, considering the latter as an additional annotator. EDAM performance was not statistically significantly different from that of six senior researchers in the field (p=0.77). The approach adopting CSO yielded the highest average agreement and also obtained the highest agreement with three out of six domain experts. Further details about this evaluation and our results are available in Osborne et al. \cite{osborne2019}.

%It was evaluated on the task of classifying papers in field of Software Architecture and its performance was not statistically significantly different from that of six senior researchers in the field (p=0.77). The approach adopting CSO yielded the highest average agreement and also obtained the highest agreement with three out of six domain experts. 

\subsection{Forecasting technology adoption}
Typically, the spreading of a technology from a one research area (e.g., Semantic Web) to a different and possibly conceptually distant area (e.g., Digital Humanities) may take several years, potentially delaying the research process.
%of technologies across research areas can result in a slow and inefficient process, since some researchers may be unaware of some relevant solutions produced by other research communities. 

The Technology-Topic Framework (TTF) \cite{osborne2017} is an approach  that suggests promising technologies to scholars in order to accelerate the pace of technology propagation. It characterises technologies according to their propagation through research topics drawn from CSO, and uses this representation to forecast the propagation of novel technologies across research fields. 

TTF was evaluated on a set of 1,118 technologies in the fields of Semantic Web and Artificial Intelligence, yielding a precision of 74.4\% and a recall of 47.7\% for the first 20 research areas. More details about Technology-Topic Framework are available in Osborne et al. \cite{osborne2017}.

%We evaluated TTF on a set of 1,118 technologies and proved to be able to forecast the adoption of these technologies in research areas such as Information Retrieval, Databases Systems, and World Wide Web. \textbf{[SE MENZIONIAMO LA VALUZATIONE METTIAMO QUALCHE NUMERO]} 

\subsection{Scientific knowledge graphs generation}
Scientific Knowledge Graphs (SKGs) are semantic graph databases that model scholarly knowledge in a structured, interlinked, and semantically rich manner. SKGs describe the actors (e.g., authors, organisations), the documents (e.g., publications, patents), and the research knowledge (e.g., research topics, tasks, technologies) in this space, as well as their reciprocal relationships.

CSO currently support two of these resources. The first is the Academia and Industry DynAmics (AIDA) knowledge graph \cite{angioni2020} whicb describes 14M papers and 8M patents in the field of Computer Science. It was generated by by automatically integrating data from Microsoft Academic Graph, Dimensions, English DBpedia, the Computer Science Ontology, and the
Global Research Identifier Database. The CSO Classifier was used to  annotate both papers and patents according to their relevant topics in CSO. 4M papers and 5M patents are also categorised according to the type of the author’s affiliations (academy, industry, or collaborative) and 66 industrial sectors (e.g., automotive, financial, energy, electronics) obtained from DBpedia. More details are available in Angioni et al. \cite{angioni2020}.
AIDA can be browsed and downloaded from \url{http://w3id.org/aida}. 

The second knowledge base integrating CSO is the Artificial Intelligence Knowledge Graph (AI-KG) \cite{dessi2020}, which is a large-scale knowledge graph that describes about 850K research entities. AI-KG includes 1,2M statements extracted from 333K research publications in the field of AI and describes 5 types of entities (e.g., tasks, methods, metrics, materials, others) linked by 27 relations. It was designed to support a large variety of intelligent services for analyzing and making sense of research dynamics, supporting researchers in their daily job, and informing decision of founding bodies and governments.

AI-KG was generated by applying an automatic pipeline that extracts entities and relationships using three tools: DyGIE++, Stanford CoreNLP, and the CSO Classifier.  It is available under CC BY 4.0 can be browsed and downloaded from \url{http://w3id.org/aikg}.

% \subsection{Academia and Industry Dynamics Knowledge Graph}
% \subsection{Artificial Intelligence Knowledge Graph}

\section{Conclusions and Future Work}
In this paper, we presented the Computer Science Ontology Framework, a conceptual framework that characterises the design, extraction and use of the Computer Science Ontology, which is currently the largest taxonomy of research topics in Computer Science. This framework includes the CSO classifier, a tool for annotating research papers according to a domain ontology. We described several high-level applications that take advantage of CSO for supporting the exploration of the research landscape and forecasting research dynamics.

%The CSO framework opens up several interesting directions of work. First, the resulting characterisation of research topics can be adopted for many kind of studies to explore several innovative solutions to explore the research landscape.

We are now working on a new version of Klink-2, in order to produce larger and more accurate ontologies of research topics.  We also plan to explore the application of this framework in other research fields, such as Engineering and Life Science.

%This will allows to extend our current technologies, such as Smart Topic Miner \cite{salatino2019b} and Smart Book Recommender \cite{thanapalasingam2018} in other domains. 

%\color{red}
%This framework primarily relies on a very large corpus of research papers. Then, we used Klink-2 to extract an ontology of research areas that allowed us to characterise the content and the topical structure of this corpus, developing the Computer Science Ontology. The natural following step was to develop the CSO Classifier that allows all the relevant stakeholders to annotate research papers according to CSO. On top of all these technologies, we can find all high-level applications that use the capabilities of the layers below.
%\color{black}

%
% ---- Bibliography ----
%
% BibTeX users should specify bibliography style 'splncs04'.
% References will then be sorted and formatted in the correct style.
%
% \bibliographystyle{splncs04}
% \bibliography{mybibliography}
%
\bibliographystyle{splncs04}
\bibliography{main}
\end{document}